\documentclass[aps,prl,twocolumn, amssymb,amsmath,floatfix,superscriptaddress]{revtex4-2}

\usepackage{dcolumn}% Align table columns on decimal point
\usepackage{bm}% bold math
\usepackage{hyperref}% add hypertext capabilities
\usepackage{physics} 
\usepackage{comment}
\usepackage{lineno}
\usepackage{braket}
\usepackage{tabularx}
\usepackage{bm}
\usepackage{euscript}
\usepackage{epsfig,psfrag}
\usepackage{graphicx}
\usepackage{color}
\usepackage{amsfonts}
\usepackage{exscale}
\usepackage{wrapfig}
\usepackage{extarrows} 
\usepackage{placeins}
\usepackage{soul}
\usepackage{amsmath}
\usepackage[english]{babel}
\usepackage[autostyle, english = american]{csquotes}
\MakeOuterQuote{"}
\usepackage{enumitem}
\usepackage{slashed} 
\usepackage{tikz}
\newcommand{\bea}{\begin{eqnarray}}  
\newcommand{\eea}{\end{eqnarray}}
\newcommand{\ben}{\begin{enumerate}}
\newcommand{\een}{\end{enumerate}}
\newcommand{\be}{\begin{equation}}
\newcommand{\ee}{\end{equation}}

\definecolor{Bcolor}{RGB}{0,0,255}
\definecolor{Rcolor}{RGB}{255,0,0}
\definecolor{Gcolor}{RGB}{0,255,0}
\definecolor{Cerulean}{RGB}{0, 123, 167}

\usetikzlibrary{external}
\begin{document}

\title{Analytically Continuing the Randomized Measurement Toolbox}

\author{Akash Vijay}
\affiliation{Department of Physics and Anthony J. Leggett Institute of Condensed Matter Theory, University of Illinois at Urbana-Champaign, Urbana, Illinois 61801, USA}
\author{Ayush Raj}
\affiliation{Department of Physics and Astronomy, Purdue University, West Lafayette, IN 47907, USA}
\author{Jonah Kudler-Flam}
\affiliation{School of Natural Sciences, Institute for Advanced Study, Princeton, NJ 08540, USA}
\author{Beno\^{\i}t Vermersch}
\affiliation{University of Grenoble Alpes, CNRS, LPMMC, 38000 Grenoble, France}
\affiliation{Quobly, 38000 Grenoble, France}
\author{Andreas Elben}
\affiliation{PSI Center for Scientific Computing, Theory and Data, Paul Scherrer Institute, 5232 Villigen PSI, Switzerland}
\affiliation{ETH Zürich - PSI Quantum Computing Hub, Paul Scherrer Institute, 5232 Villigen PSI, Switzerland}
\author{Laimei Nie}
\affiliation{Department of Physics and Astronomy, Purdue University, West Lafayette, IN 47907, USA}

\date{\today}

\begin{abstract}
We develop a framework for extracting non-polynomial analytic  functions of density matrices in randomized measurement experiments by a method of analytical continuation. 
 A central advantage of this approach, dubbed \textit{stabilized analytic continuation} (SAC), is its robustness to statistical noise arising from finite repetitions of a quantum experiment, making it well-suited to realistic quantum hardware. As a demonstration, we use SAC to estimate the von Neumann entanglement entropy of a numerically simulated quenched Néel state from Rényi entropies estimated via the randomized measurement protocol. We then apply the method to experimental Rényi data from a trapped-ion quantum simulator experiment, extracting subsystem von Neumann entropies at different evolution times. Finally, we briefly note that the SAC framework is readily generalizable to obtain other nonlinear diagnostics, such as the logarithmic negativity and Rényi relative entropies. 
\end{abstract}
                              
\maketitle

{\it Introduction.--} 
Recent advances in quantum simulator platforms enable the study of complex quantum many-body phenomena, such as thermalization, scrambling, and topological order, in well-controlled laboratory settings \cite{Preskill2018NISQ,Altman2021PRXQReview}. Understanding these phenomena requires access to the entanglement structure of quantum states. Quantum information theory provides a quantitative framework for characterizing entanglement through the von Neumann (vN) entanglement entropy of a density matrix $\rho$,
\begin{align}
   S_{vN}(\rho)=-\mathrm{Tr}(\rho\log_2\rho) ,
\end{align} which serves as a fundamental measure of bipartite entanglement \cite{horodecki2009quantum}. It plays a central role in diagnosing topological order \cite{TEE_KitaevPreskill,TEE_LevinWen}, probing thermalization and information scrambling \cite{Fast_Scramblers,ETH_EE,Deutsch_2018,Kaufman_2016,EntanglementGrowthRUCs}, and studying quantum criticality \cite{EE_Criticality,EE_PhaseTransition} and the emergence of bulk gravity in AdS/CFT~\cite{Ryu_2006,Ryu_2007}.
However, directly measuring the vN entropy for large quantum systems in quantum simulators remains experimentally challenging, motivating the search for scalable measurement approaches.

While full quantum state tomography allows complete reconstruction of a quantum state and thereby access to its vN entropy, its sample complexity scales sharply exponentially with system size, rendering it impractical beyond few-qubit systems \cite{haah2017sample,ODonnell2016Jun}.
Recent advances in randomized measurement (RM) protocols \cite{elben2019statistical, huang2020predicting,elben2023randomized} have pushed the boundary to substantially larger system sizes  for a broad class of polynomial functions of the density matrix \cite{vitale2024robust}. Such functions can also be estimated utilizing many-body interference between multiple physical copies of quantum state prepared simultaneously in a quantum experiment \cite{islam2015measuring, Kaufman_2016}.
These protocols enable measurement of scalable quantities such as Rényi entropies $S_{k}(\rho)=\frac{1}{1-k}\log_2\mathrm{Tr} \rho^{k}$, $k\geq 2$ integer \cite{elben2018renyi,elben2019statistical}, Hilbert–Schmidt fidelities, and mixed-state entanglement witnesses \cite{elben2020mixed}. Such quantities serve as powerful diagnostics of correlations and entanglement \cite{brydges2019probing,elben2020mixed}, but they are not entanglement measures \cite{horodecki2009quantum}. Moreover, Rényi entropies can display qualitatively different behavior from the vN entropy~\cite{tibor2019diffusion,wang2025singularity,bertini2022growth,agon2023renyi}.  Therefore, reliable estimates of the vN entropy are essential for a quantitative characterization of entanglement in many-body quantum systems.

\begin{figure}[t]
    \includegraphics[width = \columnwidth]{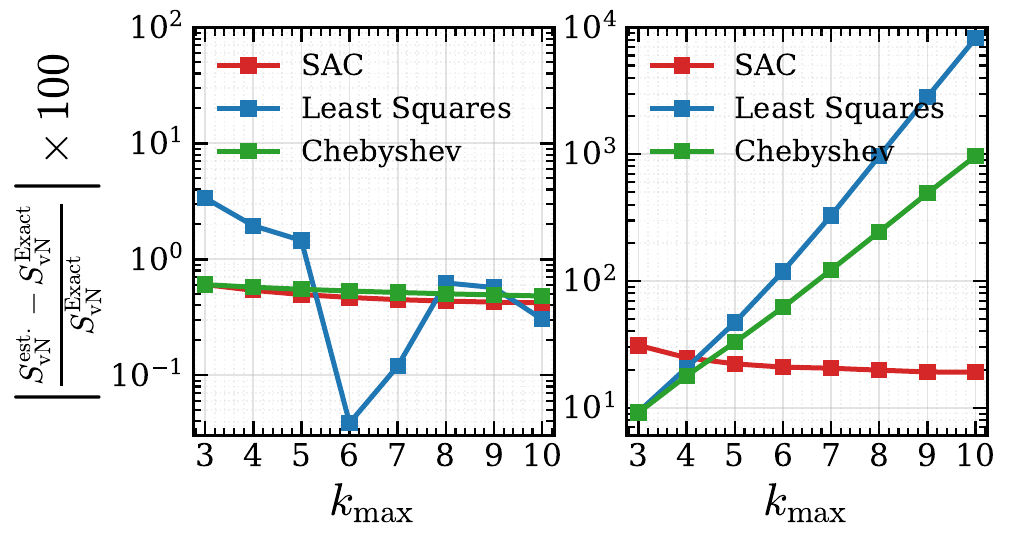}
    \caption{SAC and traditional fitting methods~\cite{vermersch2024enhanced, ChebyshevBook} estimating the von Neumann entropy of an $L = 7$ subsystem in the ground state of the transverse-field Ising model $H=-J\sum_i\sigma_i^z\sigma_{i+1}^z-h\sum_i\sigma_i^x$ with periodic boundary conditions, away from criticality ($J=1$, $h=0.5$, $L_{\text{total}}=15$).  Error percentages of different methods are shown as a function of the largest R\'enyi index in the dataset.  Left: for noiseless R\'enyi inputs. Right: for inputs with 10\% independent Gaussian noise, averaged over 200 realizations.}
    \label{fig:TFIM_Numerics}
\end{figure}

In this Letter, we develop a framework to reliably estimate the vN entropy in quantum simulation experiments using a finite number of measured Rényi entropies. While the vN entropy can, in principle, be obtained from knowledge of all integer Rényi entropies via analytic continuation,
$S_{vN}=\lim_{k\to1^{+}}S_{k}$,
a consequence of Carlson’s theorem~\cite{hardy1920two}, in practice only a limited number of Rényi orders are accessible, each with finite statistical uncertainty. This renders the continuation problem ill-posed and highly sensitive to statistical noise—much like the analytic continuation of Green’s functions away from the
Matsubara domain or extracting physical scattering amplitudes in quantum chromodynamics~\cite{Maximum_Entropy_Method,Bayesian_Inference,SVZ_SumRules,Sum_Rules_Review,Smearing_Method}. We employ the \emph{ stabilized analytic continuation} (SAC) approach of Ciulli and Spearman~\cite{CS1,CS2,CS3,CS4,CS5}, which reformulates analytic continuation as a constrained convex optimization problem: among all analytic functions compatible with the data, SAC selects the one minimizing a chosen $L^{2}$-norm, which measures the amount of structure induced by the continuation near the target point. Our novel adaptation of SAC yields stable and noise-resilient estimates of the vN entropy even when only a few Rényi entropies are available. As a first controlled illustration, Fig.~\ref{fig:TFIM_Numerics} shows the performance of SAC in an example of estimating the half-chain vN entropy of a spin chain model, 
and compares it against more traditional extrapolation schemes~\footnote{We also tested Padé-type rational approximants, but found them  to be sensitive to noise and rational-order choices, and in some cases unable to return a physical estimate. See also discussion around Fig.~\ref{fig:simulation}}. Here independent Gaussian noise is used as a preliminary test of noise sensitivity and is also relevant when different R\'enyi entropies are estimated in independent experiments, such as many-body interference protocols \cite{islam2015measuring, Kaufman_2016}. The correlated-noise setting relevant to randomized measurements is treated in the simulation benchmark below using the covariance-aware SAC protocol.
Beyond entanglement entropy, our framework applies broadly to estimating general non-polynomial spectral functions of quantum states accessible through analytic continuation.

Next, we present our adaptation of the SAC method, including deriving an explicit analytic expression of the vN entropy estimate from a few Rényi entropies, and generalizing to the case with (correlated) statistical noise. 
We benchmark our framework using numerically simulated Rényi data for a quenched Néel state and demonstrate its performance on experimental data from a trapped-ion quantum simulator~\cite{brydges2019probing}, where it enables extraction of subsystem vN entropies across time.

{\it Stabilized analytic continuation.--}\label{sec:Stabilized continuation main text}
Let $S_{z}(\rho)=\frac{1}{1-z}\log_2\mathrm{Tr}\rho^{z}$, $z \in \mathbb{C}$, denote the R\'enyi function on the complex plane. Suppose we are provided noisy estimates of this function at $N$ integer points $z = 2,3,.\cdots,k_{\text{max}}$.
Our task is to analytically continue this dataset to the vN point, $z = 1$. 
Assuming the true underlying function is smooth, conventional approaches typically involve polynomial fitting. However, the extrapolated value can depend on several modeling choices, including the functional form of the fit (e.g. $z$ vs. $1/z$), the polynomial degree, and the basis used to represent it. Lower-degree fits may reduce variance in the presence of noise but can introduce bias, while higher-degree polynomials can suffer from overfitting~\cite{StatLearningBook,mohammadipour2025ZNE, Bishop2006MachineLearning}.

\begin{figure}[h]
 \includegraphics[width = \columnwidth]{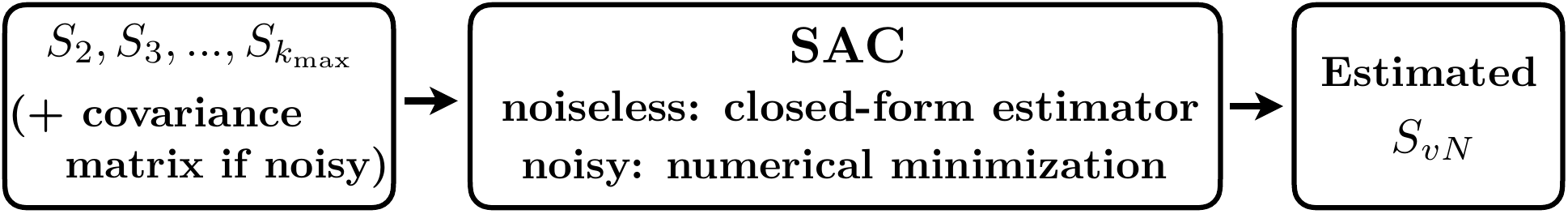}
    \caption{SAC protocol for estimating the von Neumann entropy from finite Rényi inputs, with correlated noise encoded through the covariance matrix.}
    \label{fig:flowchart}
\end{figure}

\begin{figure*}[t]
    \centering
    \includegraphics[width=0.95\linewidth]{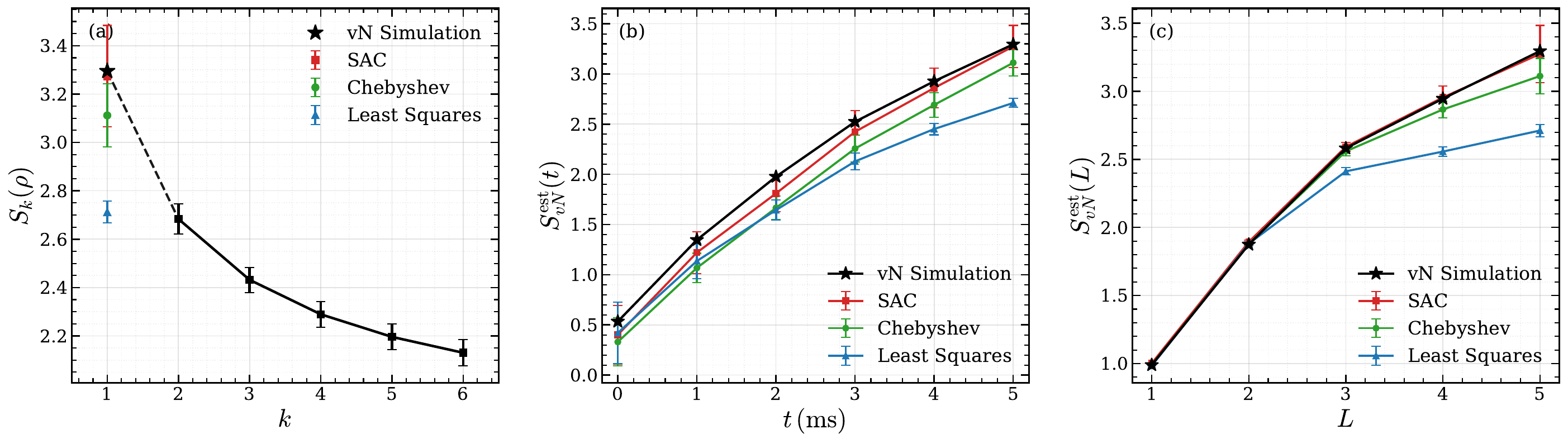}
    \caption{Numerical benchmarks using simulated data for a 10-qubit N\'eel state quenched under Hamiltonian~\eqref{eq:XY Hamiltonian} with decoherence included. Left: von Neumann entropy estimates from different methods at fixed time, $t=5$ms, and subsystem size, $L=5$ sites. Center: von Neumann entropy versus time for $L= 5$.
Right: von Neumann entropy versus subsystem size at $t = 5$ms.}
    \label{fig:simulation}
\end{figure*}

This leads us to the method of stabilized analytic continuation (SAC), which bypasses these challenges by exploiting the fundamental analytic properties of the underlying function~\cite{CS1,CS2,CS3,CS4,CS5}. To adapt it to the vN entropy problem, we construct a variational function whose minimum $L^2$-norm determines the optimal estimate of the vN entropy. We explicitly derive the analytical expression for the vN estimate resulting from this variational approach in the noiseless case. We also extend the SAC framework, originally formulated for uncorrelated noise, to handle correlated noise. Our adaptation is found to be more robust to realistic noise than conventional extrapolation techniques, as confirmed using data with correlated noise below. Fig.~\ref{fig:flowchart} summarizes the workflow.

In the first part of the analysis, we assume the data points are noiseless. We begin by defining the following {\it discrepancy function} on the $z$ plane
\begin{align}\label{eq:Discrepancy function}
    D_{\alpha}(z) = \frac{S_{z}(\rho)}{z-1} - \frac{\alpha}{z-1}
\end{align}
where $\alpha$ is a real variational parameter. 
$D_{\alpha}(z)$ serves as an auxiliary variational construction: it assumes the values $\{d_{i}(\alpha) = \frac{S_{i}(\rho)}{z_{i}-1} - \frac{\alpha}{z_{i}-1} \}$ at the points $\{z_{i} = 2,\cdots, k_{\text{max}}\}$, and its $L^2$-norm will be minimized over analytic continuations compatible with these data and by varying $\alpha$. By dividing $S_{z}(\rho)$ by $(z-1)$, we artificially introduce a simple pole at the vN point, $z=1$. 
The residue of $D_{\alpha}(z)$ at this pole is $S_{vN}(\rho) - \alpha$. Thus, when $\alpha = S_{vN}(\rho)$, the pole is canceled, which motivates using the norm minimization to select the optimal $\alpha$.

We now map the analytic domain of $D_{\alpha}(z)$ to the interior of the unit disk via a conformal transformation $w(z)$, chosen such that 
\begin{enumerate}
    \item $D_{\alpha}(z(w))$ remains analytic for $|w|<1$
    \item The real half-line $[1,\infty)\in \mathbb{R}_{z}$ is mapped to the interval $[-1,1] \in \mathbb{R}_{w}$
    \item The point $z=1$ is mapped to $w=-1$. 
\end{enumerate}
To construct such a map, one must first identify the domain of analyticity of the Rényi function $S_{z}(\rho)$ in the complex $z$ plane. In general, this domain depends both on the dimension of the Hilbert space as well as the spectrum of $\rho$. Using a perturbative argument, one can show that $S_z(\rho)$ remains analytic on a semi-infinite strip with $\Re z > 1$ and $|\Im z| < \frac{c}{\log(d)}$, where $d$ is the dimension of the Hilbert space and $c \sim \mathcal{O}(1)$. One caveat is when this analytic strip is very narrow, either at very large $d$ or small $c$; in that case the conformal map compresses the integer R\'enyi points near the boundary of the unit disk, reducing the numerical stability and reliability of the SAC estimate. In practice, however, direct numerical tests on randomly sampled density matrices indicate that the first branch points of $S_z(\rho)$ typically appear much further away than this conservative bound. A generic semi-infinite strip of half-width $\pi \epsilon/2$ centered on the real axis, and with $\Re z>1$, can be conformally mapped to the unit disk via a two-step conformal transformation: 
\begin{equation}
\label{eq:conformalmapping}
 \xi = \cosh(\frac{z-1}{\epsilon} + \frac{i \pi}{2}), \mbox{ followed by } \hspace{5pt}  w = \frac{\xi-\eta i}{\xi+ \eta i}
\end{equation}
Here $\eta$ is a free parameter. In practice, $\eta$ and $\epsilon$ are chosen from simulation-data benchmarks, corresponding to the regime where the pole contribution dominates the norm in~\eqref{eq:Norm}. The Rényi function in the unit disk, $S_w \equiv S_{z(w)}$, takes the values ${S_2(\rho),\cdots,S_{k_{\text{max}}}(\rho)}$ at the mapped points $w_i=w(i)$, for $i=2,\cdots,k_{\text{max}}$. Since $S_w$ is analytic for $|w|<1$, so is $D_\alpha(w)\equiv D_\alpha(z(w))$. However, $D_\alpha(w)$ retains a pole on the unit circle at the mapped vN point, $w=-1$. We quantify the structure of the function on the unit circle through the following $L^2$ (pseudo-) norm~\footnote{We remark here that the $L^{2}$ (pseudo-)norm defined in \eqref{eq:Norm} is not the only choice one can make. In fact, there are four different $L^{2}$ type norms one can write down where the integrand is expressed as the square of the real part of the function, the imaginary part of the function, the angular derivative of the real part of the function or the angular derivative of the imaginary part of the function. The choice ~\eqref{eq:Norm} is motivated by the observation that the pole at $w=-1$ only manifests as a divergence in the imaginary part of the function.}: 
\begin{align}\label{eq:Norm}
    ||G|| = \frac{1}{2 \pi }\int_{0}^{2\pi} \bigg|\frac{d \text{Im} G(e^{i\theta})}{d\theta} \bigg|^{2} d\theta
\end{align}
Note that $||G|| = 0$ does not imply $G = 0$; it only implies $G(w)$ must be a constant in the interior of the disk.
We can restrict to the set of functions which vanish at some point $w_{0}$ in the disk, then (\ref{eq:Norm}) defines a valid norm. Often, we choose $w_{0} = w_{2}$, the location of the first input data point, and replace $D_\alpha(w)$ with $D^\prime_\alpha(w) = D_\alpha(w) - d_2(\alpha)$. 

Crucially, if $\alpha \neq S_{vN}(\rho)$, $|| D^\prime_\alpha(w)||$ will be very large due to the presence of the singularity on the unit circle. %(formally diverges). 
We can therefore estimate the optimal $\alpha_{\min} = S_{vN}(\rho)$ by carrying out the following dual minimization steps:

\begin{enumerate}
    \item For fixed $\alpha$, we search for the analytic function $Y_{\alpha}(w)$ which takes the desired values $\{ Y_{\alpha}(w_i) = D^\prime_{\alpha}(w_{i}) = d_{i}(\alpha) - d_{2}(\alpha)\}$ at the points $\{ w_{i}\}$ and minimizes the norm $||Y_{\alpha}||$. This minimization can be recast into a linear matrix optimization problem by leveraging the analyticity of $Y_{\alpha}(w)$~\cite{CS1}.
    The minimal norm $$\delta(\alpha) \equiv \min_{Y_{\alpha}(w_i) = D^\prime_{\alpha}(w_{i})}{|| Y_{\alpha}||}$$ provides a measure of the structure that is forced on the function by the data itself. 
    \item Next we minimize $\delta(\alpha)$ over all possible values of $\alpha$. This is where we expect the pole at $w=-1$ cancels out, and the minimal value $\alpha_{\min}$ serves as our best estimate for $S_{vN}(\rho)$. 
\end{enumerate}

%In the noiseless case, 
Due to the absence of noise, both of these minimization steps can be performed exactly to obtain the following closed-form expression for the vN estimate:
\begin{align}
\label{eq:alphamin}
    S^{\text{est}}_{vN} = \frac{\sum_{i,j =3}^{k_{\text{max}}}(A^{-1})_{ij}\Bigl(\frac{S_{i}(\rho)}{i-1} -S_{2}(\rho)\Bigl)\Bigl(\frac{1}{j-1} - 1\Bigl)}{\sum_{i,j =3 }^{k_{\text{max}}}(A^{-1})_{ij}\Bigl(\frac{1}{i-1} - 1\Bigl)\Bigl(\frac{1}{j-1} - 1\Bigl)}
\end{align}
where $A_{ij}$ is a symmetric and positive definite matrix (refer to~\cite{SM} for its expression).

In the presence of noise, Step 1 above is modified to include sampling of all data points that lie within a suitable neighborhood of the mean values $D^\prime_{\alpha}(w_j)$. If $C$ denotes the covariance matrix of the data points, then the goodness of fit of an arbitrary point $\{y_{j}\}$ in data space can be measured using a $\chi^{2}$ statistic: $\chi^2(\mathbf{y};\alpha) = \sum_{i,j=2}^{k_{\text{max}}} (y_i - D^\prime_{\alpha}(w_i)) (C'^{-1})_{ij} (y_j - D^\prime_{\alpha}(w_j))$, where $\mathbf{y} = \{y_2, ..., y_{k_{\text{max}}} \}$, and $C'_{ij} =\frac{C_{ij}}{(z_i-1)(z_j-1)}$ is the covariance matrix of the corresponding discrepancy function. Note that for the noisy data case, we select an arbitrary subtraction point on $(-1,1)$ and treat its value as an additional variational parameter~\cite{CS2, CS5}, and hence the sum runs from $i,j = 2$. The minimal norm $\delta(\alpha)$ is now computed by minimizing over all possible points $\textbf{y}$ subject to the constraint $\chi^2(\mathbf{y}; \alpha) \leq \chi^2_0$ for some constant $\chi^2_0$. Namely,
\begin{equation}
    \delta(\alpha) \equiv \min_{\chi^2(\mathbf{y};\alpha) \leq \chi^2_0}{|| Y_{\alpha}||} 
\end{equation}
Step 2 of the minimization procedure stays the same. In this case, we are unable to provide a closed-form expression for $S^{\text{est}}_{vN}(\rho)$, but we can still efficiently estimate this value numerically for arbitrarily large systems~\cite{SM}, where $\chi^2_0$ is chosen such that the numerical minimizations converge.

{\it Benchmarking SAC with randomized-measurement simulation data-}
To benchmark SAC in an experimentally relevant setting, we generate on a classical computer the input Rényi entropies using the standard randomized-measurement (RM) and classical-shadow protocol. For a system of $L$ qubits, local random unitaries 
$U=U_1\otimes\cdots\otimes U_L$ are drawn independently from a unitary 3-design~\cite{elben2023randomized} and followed by projective measurements in the computational basis.  From $N_u$ random unitaries and $N_m$ repetition measurements per unitary, one constructs classical shadows, from which polynomial functions of the density matrix such as
trace moments can be estimated via U-statistics~\cite{elben2020mixed}. Using these trace moments, one can then obtain R\'enyi entropies. Non-polynomial functions of the density matrix, such as vN entropy, are not directly accessible within the RM framework. In practice, RM can provide up to $k_{\text{max}} = 6$ orders of Rényis using a batch shadow formalism~\cite{rathentanglement2023}; higher Rényis suffer from increasing statistical uncertainty. These low-order Rényi estimates and their covariance matrix serve as input to the noisy SAC procedure.

Our simulation is performed on a 10-qubit system initialized in a low-entropy Néel state and subsequently evolved under a spin-$1/2$ Hamiltonian~\cite{brydges2019probing,Porras2004PRL},
\begin{equation}\label{eq:XY Hamiltonian}
H=\sum_{i<j}J_{ij}(\sigma_i^+\sigma_j^-+\sigma_i^-\sigma_j^+)+ B\sum_j\sigma_j^z,
\end{equation}
where $\sigma_i^{\pm}$ are spin ladder operators and the couplings follow an approximate power-law decay $J_{ij} \approx J/|i-j|^{1.2}$ with $J=420\mathrm{s}^{-1} \ll B$. The simulations incorporate realistic decoherence sources, including imperfect Néel-state preparation, spontaneous emission, spin flips, and depolarization from local unitaries~\cite{brydges2019probing}.

RM are simulated on a classical computer to construct classical (batch) shadows, which are then used to estimate Rényi entropies~\cite{elben2020mixed}. 
We perform 1000 independent shadow experiments, each consisting of $N_u = 500$ random unitaries and $N_m = 150$ measurements per unitary. Using U-statistics, Rényi entropies of orders $k = 2,\cdots,6$ are computed for each experiment. The vN entropy is extracted with both SAC and traditional fitting schemes, including Chebyshev polynomial interpolation~\cite{ChebyshevBook} and least-squares fitting~\cite{vermersch2024enhanced} \footnote{As mentioned before, we also tested Padé-type rational approximants as an additional baseline. While they can perform reasonably well for noiseless data, we found them unreliable for noisy randomized-measurement data: the results depend sensitively on the rational order and on filters (for example, excluding spurious poles near $n=1$). In particular, once these filters are imposed, the covariance-aware noisy rational fit does not consistently return an accepted physical estimate. We therefore do not include rational fits as a main benchmark.}. These estimates are compared with the exact vN entropy obtained directly from the simulation (Fig.~\ref{fig:simulation}). 
For each of the 1000 independent shadow datasets, we treat the Rényi entropies as an independent input instance and obtain a separate vN entropy estimate from each continuation method. For noisy SAC, the Rényi covariance matrix is estimated for each dataset by jackknife resampling the shadow density matrices. The resulting mean Rényi vector and jackknife covariance matrix are then supplied to SAC, and the final estimate and statistical uncertainty are obtained from the distribution of SAC outputs over the 1000 datasets. The same independent-dataset analysis is applied to Chebyshev extrapolation and least-squares fitting.

{\it Application to Experimental Data from a Trapped-ion Quantum Simulator--}
Finally, we apply SAC to extract the vN entropy from Rényi data obtained in a trapped-ion experiment. Unlike the simulated benchmark of Fig~\ref{fig:simulation}, where exact ground truth allows a quantitative comparison among extrapolation methods, the trapped-ion results here are intended as a proof-of-principle demonstration on a real device. The system contains ten $^{40}\text{Ca}^+$ ions, prepared in an approximate N\'eel state and evolved under Hamiltonian~\eqref{eq:XY Hamiltonian}.

\begin{figure}[t]
    \centering
    \includegraphics[width=1
    \columnwidth]{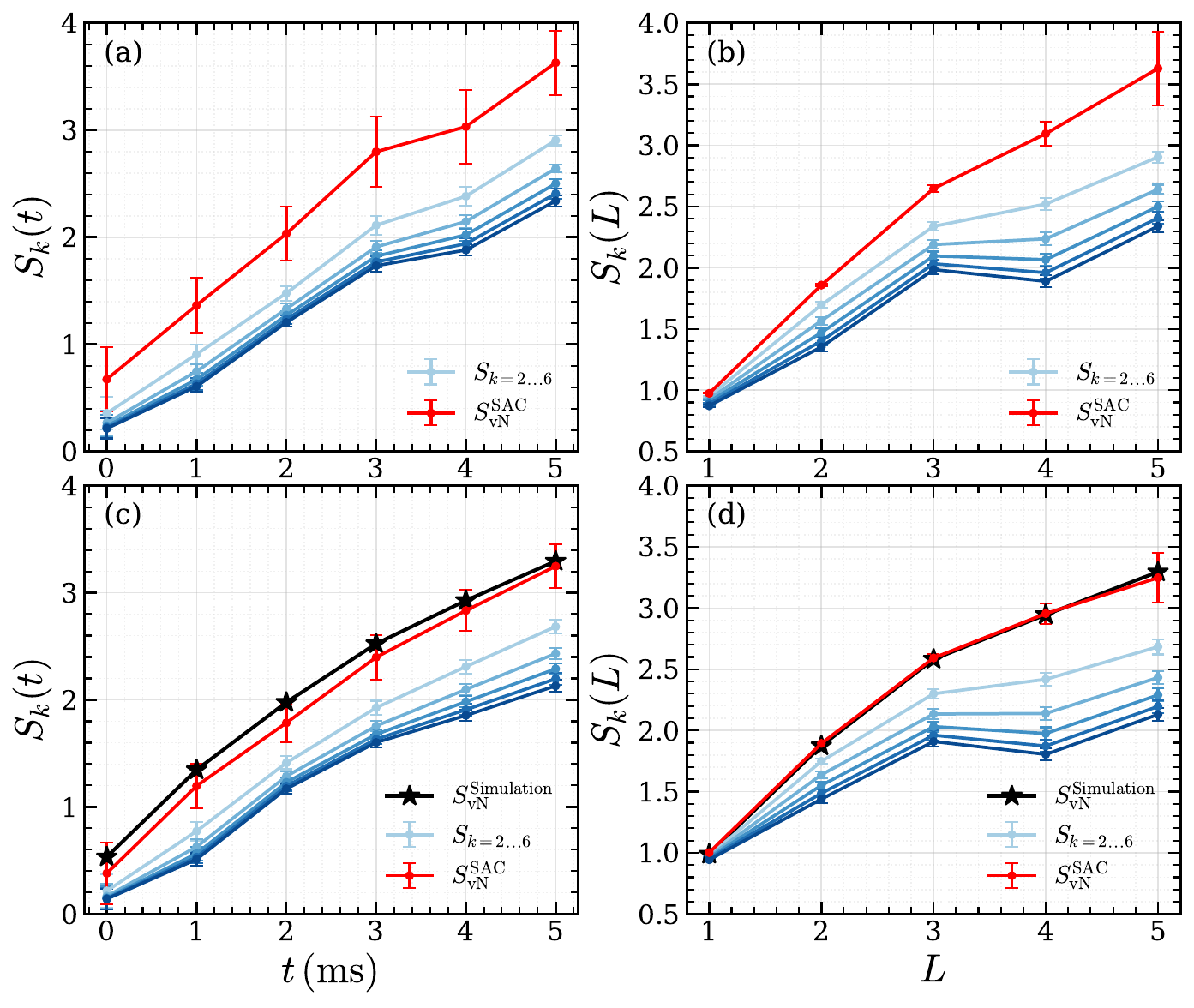}
    \caption{(a)(b): $S_{vN}$ as a function of time $t$ and subsystem size $L$ obtained via SAC in a trapped-ion experiment with ten $^{40}\text{Ca}^+$ ions, initialized in an approximate N\'eel state and evolved under Hamiltonian~\eqref{eq:XY Hamiltonian}. 
   For comparison, $S_{vN}$ obtained via SAC using simulation data, as well as R\'enyi data from simulated randomized measurements, are plotted in (c)(d). 
     In (a)(c), $L = 5$. In (b)(d), $t = 5$ms.}
    \label{fig:experiment}
\end{figure}

The experimental dataset comprises a single shadow tomography experiment with $N_u = 500$ and $N_m = 150$ for various timesteps and subsystem sizes. Jackknife resampling is applied to obtain the mean Rényi entropies for orders $k = 2,\cdots,6$ and their covariance matrix.\footnote{The mean and covariance matrix are jackknife-corrected.} While direct jackknife resampling would naively require $\mathcal{O}(N_B)$ evaluations of the U-statistic, where $N_B$ is the number of batch shadows, we employ a fast implementation based on hashing. The resulting mean Rényi vector and covariance matrix are used as input to SAC, yielding the vN entropy estimates shown in Fig.~\ref{fig:experiment}. The SAC estimator variance, whose square root gives the error bar in Fig.~\ref{fig:experiment}, is estimated using a double-jackknife protocol~\cite{SM}.

{\it Discussion--}
We introduced a robust framework for extracting the vN entropy from finite sets of Rényi entropies subject to statistical noise obtained via randomized measurements. When tested on controlled examples, including the synthetic benchmarks of Fig.~\ref{fig:TFIM_Numerics} and the randomized-measurement simulations of Fig.~\ref{fig:simulation}, SAC yields consistently stable estimates, whereas polynomial-based extrapolations can be more model-dependent and more sensitive to noise. We further presented a proof-of-principle application to experimental data from a trapped-ion quantum simulator following a quantum quench. Since randomized measurements are now routinely used to measure entropies in many platforms, our noise-resilient approach provides a practical post-processing tool for analyzing data affected by statistical uncertainty and may substantially reduce measurement overheads. It also extends to situations where R\'enyi entropies are measured through many-body interference experiments \cite{islam2015measuring,Kaufman_2016}.

There are many potential applications of SAC beyond extracting vN entropies. The method naturally extends to any information-theoretic quantity accessible through a replica trick—such as logarithmic negativity, Petz–Rényi divergences, or Uhlmann fidelity~\cite{elben2020mixed,lieb1973convex,uhlmann1977relative, petz1986quasi, Lennert2013RenyiEntropies, Wilde2014sandwiched,  Datta2015alphaz,KudlerFlam2024RMI}—which we leave for future exploration. More broadly, SAC offers a promising strategy for addressing long-standing analytic-continuation challenges in many other areas of physics, from quantum Monte Carlo studies to the reconstruction of scattering amplitudes.

{\it Acknowledgement--} We thank Hsin-Yuan Huang, Nima Lashkari, Jong Yeon Lee, Eric Schultz, Vandana Ramakrishnan, and Daniel Mark for helpful discussions.
JKF is supported by the Marvin L.~Goldberger Member Fund at the Institute for Advanced Study and the National Science Foundation under Grant No.~PHY-2514611. 
BV acknowledges support from the hybrid quantum
initiative (HQI ANR-22-PNCQ-0002).

Note: The open-source code to extract R\'enyi entropies from batch shadows, applied to the experimental data of Ref.~\cite{brydges2019probing}, is available on \href{https://github.com/bvermersch/RandomMeas.jl/blob/main/examples/BrydgesScience2019.ipynb}{GitHub}, and presented in details in Ref.~\cite{elben2025randommeasjl}. The numerical simulation data used in this work can be made available upon request.

\bibliography{References}% 

\appendix

\onecolumngrid

\section{Appendix A: Stabilized Analytic Continuation}\label{sec:stabilized_analytic_continuation_detailed}

In this Supplemental Material, we review the key features of the SAC method outlined in references ~\cite{CS1,CS2,CS3,CS4,CS5}. Additional details can be found in the original references.

The basic setup of the problem is as described in the main text. $F(w)$ is an unknown analytic function that is holomorphic in the interior of the unit disk. It assumes real values on the real axis: $F(w_i) = a_i \in \mathbb{R}$ for $w_i \in (-1,1)$ and $i = 2,3,..., N+1$, where $N$ denotes the number of data points ($N = k_{\text{max}} - 1$ in the main text).  
Our aim is to fit for the value of the function at a specific point on the unit circle, which we take to be $w=-1$ without loss of generality. SAC was originally developed by Ciulli and Spearman~\cite{CS1,CS2,CS3,CS4,CS5} to diagnose resonances 
of the Green's function which lie outside, but close to the unit disk, using a finite set of noisy data points provided within the interior of the disk. The presence of nearby poles results in a significant amount of structure in the function on the unit circle, which can be measured by a suitable choice of norm. Minimizing this norm over all consistent analytic continuations then yields the minimal amount of structure that is forced upon the function by the data itself.

To adapt this method to our target problem, we first artificially introduce a pole at $w=-1$ by defining $G(w) = \frac{F(w)}{w+1}$. The aim now is to estimate the residue of the pole at $w=-1$. To do so, as explained in the main text, we define the discrepancy function $D_{\alpha}(w) = G(w) - \frac{\alpha}{w+1}$ where $\alpha$ is a variational parameter~\footnote{
The discrepancy function $D_{\alpha}(w)$ used in this appendix differs slightly from the convention used in the main text. Here, to keep it general, we formulate the SAC construction directly in the $w$ plane and introduce the pole as $1/(w+1)$. In the main text, the discrepancy is instead defined first in the original $z$ plane as
$D_{\alpha}(z)=[S_z(\rho)-\alpha]/(z-1)$,
and then mapped to the disk, i.e. $D_{\alpha}(w)\equiv D_{\alpha}(z(w))$. The two conventions are related by the known conformal map $z=z(w)$. Both conventions encode the same pole-cancellation condition at the von Neumann point.
}. 
If $\alpha= F(-1)$, the residue of $D_\alpha(w)$ is 0 at $w=-1$, hence $D_\alpha (w)$ becomes analytic even on the unit circle. To measure the structure of the function near $w=-1$, we introduce a norm defined entirely in terms of the values of the function on the unit circle. Given either a Dirichlet or a Neumann boundary problem, there are two corresponding unique norm choices. Both are continuous $L^{2}$-type norms which stem from an inner product. For the problem of analytic continuation of R\'enyi entropies, a proper choice of the norm is
\begin{equation}\label{eq:norm choice}
    ||X|| = \frac{1}{2 \pi}  \int_0^{2 \pi} \bigg|\frac{\partial}{\partial \theta}\text{Im}X(e^{i \theta})\bigg|^2 d\theta = \frac{1}{2\pi}  \int_0^{2\pi} \bigg|\frac{\partial}{\partial r}\text{Re}X(re^{i \theta})\big|_{r = 1}\bigg|^2 d\theta
\end{equation}
where the latter equality follows from the Cauchy-Riemann equations. Importantly, this norm is singular when the function $X$ possesses a pole at $w = -1$.  
For the case of a constant non-zero function $X$, $||X||$ is zero and is a pseudo-norm. To rule out this special case, we restrict $X$ to be the subset of analytic functions with $X(w_{0})=0$, where $w_0$ is an arbitrary subtraction point. In what follows, we replace $D_{\alpha}(w)$ with $D'_{\alpha}(w) = D_{\alpha}(w)- D_{\alpha}(w_0)$. A convenient choice for $w_{0}$ is $w_{2}$, the location of the first data point. 
Let us introduce the notation
\begin{align}
    x_{,r}(\theta) = \frac{\partial \text{Re}X(re^{i\theta})}{\partial r}\bigg|_{r=1} 
\end{align}
As illustrated in \cite{CS1,CS2,CS3,CS4,CS5}, knowing $x_{,r}(\theta)$ on the unit circle allows one to reconstruct the function $X(w)$ in the interior of the unit disk upto an overall constant through the formula
\begin{equation}
    X(w) = X(w') - \frac{1}{\pi} \int_0^{2\pi} x_{,r}(\theta) \ln\Bigg (\frac{e^{i \theta} - w}{e^{i \theta} + w'}\Bigg) d \theta
\end{equation}
Furthermore, if $w'$ is chosen to be the subtraction point $w_0$,
then $x_{,r}(\theta)$ fully determines $X(w)$ in the interior of the disk. 

In our case, $X(w) = D'_{\alpha}(w)$, the subtracted discrepancy function. $D'_{\alpha}(w)$ is required to take on the values 
\begin{align}
  d'_{\alpha;i} 
  = \bigg( \frac{a_{i} }{w_{i}+1} - \frac{a_{2} } {w_{2}+1}\bigg) - \bigg( \frac{\alpha }{w_{i}+1} - \frac{\alpha}{w_{2}+1}\bigg) 
\end{align} 
at the data points $\{w_{i}\}$.  We have chosen the first data point to be the subtraction point, i.e. $w_0 = w_2$. There is an infinite family of analytic functions $D'_{\alpha}(w)$ that satisfy the condition.
From this infinite family, we need to select the analytic continuation that has the smallest norm. As illustrated in  ~\cite{CS1,CS2,CS3,CS4,CS5}, this non-linear minimization problem can be recast into a linear matrix optimization problem, and the minimal norm (squared) $\delta_{\min}^{2}$ of $D'_\alpha(w)$ is given by 
\begin{equation}
\label{eq:delta2min}
    \delta_{\min}^{2} =  \sum_{i,j = 3}^{N+1}  (A^{-1})_{ij}   d'_{\alpha;i}d'_{\alpha;j}
\end{equation}
Here the sum starts from $i,j =3$ because we have chosen the subtraction point to be the first data point $w_0 = w_2$. $A$ is an $(N-1)\times (N-1)$ positive-definite, symmetric matrix with matrix elements defined as
\begin{align}
     A_{ij} = \frac{2}{\pi} \int_0^{2 \pi} \ln\Bigg |\frac{e^{i \theta} - w_i}{e^{i \theta} - w_2}\Bigg| \ln\Bigg |\frac{e^{i \theta} - w_j}{e^{i \theta} - w_2}\Bigg|    d\theta
\end{align}
Note $A_{ij}$ only depends on the data points (locations) $w_{i}$ and not the data values themselves. $\delta_{\min}^{2}$ still depends on $\alpha$ and as we vary $\alpha$, it should have a sharp minimum when $\alpha$ equals the true residue of $G(w)$, namely $\alpha = F(-1)$. $\delta_{\min}^{2}$ turns out to be a quadratic function of $\alpha$, and we can immediately write down a closed form expression for $\alpha$ that minimizes this quadratic function:
\begin{align}
    \alpha_{\min} = \frac{\sum_{i,j =3}^{N+1}(A^{-1})_{ij}\Bigl(\frac{a_{i}}{w_{i}+1} - \frac{a_{2}}{w_{2}+1}\Bigl)\Bigl(\frac{1}{w_{j}+1} - \frac{1}{w_{2}+1}\Bigl)}{\sum_{i,j =3 }^{N+1}(A^{-1})_{ij}\Bigl(\frac{1}{w_{i}+1} - \frac{1}{w_{2}+1}\Bigl)\Bigl(\frac{1}{w_{j}+1} - \frac{1}{w_{2}+1}\Bigl)}
\end{align}
This is our estimate for the target value $F(-1)$~\footnote{ Note that this expression is slightly different from Eq (5) in the main text, because here to keep it general we have chosen to directly work in the $w$ plane.}.

\section{Appendix B: Stabilized Analytic Continuation with Errors}\label{sec:Noisy Stabilized Continuation uncorrelated}
We now consider the case where the prescribed data values $\{ d_{i}\equiv d'_{\alpha;i}\}$ are noisy. For notational simplicity, we shift the indexing from $(2,3,\cdots ,N+1)$ to $(1,2,\cdots , N)    $. The associated errors are generically correlated and can be described by a symmetric, positive semi-definite covariance matrix $C$: 
\begin{align}
    C = \sum_{I=1}^{N}\epsilon^{2}_{I} {\bf e}_{I} {\bf e}_{I}^{T}
\end{align} 
where ${\bf e}_{I}$ denote the orthonormal eigenvectors and $\epsilon^{2}_{I} \geq 0$ denote the non-negative eigenvalues. Equivalently, we may write $C = O D O^{T}$ where $D = \text{diag}(\epsilon^{2}_{1}, \cdots, \epsilon^{2}_{N})$ and $O = ({\bf e}_1,\cdots, {\bf e}_{N})^{T}$

The $N$ data points are now viewed as a point in $\mathbb{R}^{N}$, denoted by $\mathbf{d} = (d_{1},d_{2},\cdots, d_{N})^T$. An arbitrary point in this data space is denoted as $\mathbf{y} = (y_{1},y_{2},\cdots, y_{N})^T$.  The goodness of fit of an arbitrary data point with the specified data values is measured by a $\chi^{2}$ statistic: 
\begin{align}
    \chi^{2} = (\mathbf{y} - \mathbf{d})^{T}C^{-1}(\mathbf{y} - \mathbf{d}) =\sum_{I=1}^{N}\frac{(\tilde{y}_{I} - \tilde{d}_{I})^{2}}{\epsilon_{I}^{2}}
\end{align}
In the second equality, we have expanded both $\mathbf{d} = \sum_{I=1}^{N} \tilde{d}_{I}{\bf e}_{I}$ and $\mathbf{y} = \sum_{I=1}^{N} \tilde{y}_{I}{\bf e}_{I}$ in the eigenbasis ${\bf e}_{I}$ of the covariance matrix $C$.
We shall restrict ourselves to the set of points which satisfy $\chi^{2}\leq \chi^{2}_{0}$ for some chosen constant $\chi^{2}_{0}$. These points constitute an ellipsoid in data space whose principal axes correspond to ${\bf e}_{I}$. 

Due to errors in the data points, it now no longer makes sense to choose the subtracted value to correspond to one of the data values. To incorporate errors,  we fix the subtraction location to the first input point, $w_0 = w_2$, while treating the corresponding function value as an additional variational parameter~\cite{CS2, CS5}. We have checked that different choices of subtraction point locations give indistinguishable estimates. Given any data point $\mathbf{y}$ within this ellipse $\chi^{2}\leq \chi^{2}_{0}$, the minimal (squared) norm $\delta^{2}_{\min}$ is given by
\begin{align}
   \delta^{2}_{\min} = (\mathbf{y} - \mathbf{y}_{0})^{T}A^{-1}(\mathbf{y} - \mathbf{y}_{0}) 
\end{align}
Here $\mathbf{y}_{0} = y_{0}\mathbf{1}$ where $\mathbf{1} = (1,1,\cdots,1)^T$ and $y_{0}$ is the value of the guess function at $w_0$. Like $w_0$, $y_0$ is also a free parameter. 
Again it is convenient to express this quantity in the ${\bf e}_{I}$ basis. To this end, let us first expand the $\mathbf{1}$ vector as $\mathbf{1} = \sum_{I=1}^{N}1_{I} {\bf e}_{I}$ where $1_{I} = {\bf e}_{I}^{T}\bf{1}$. Next, we define a new matrix $B = O^{T}AO$. Due to the orthogonality of $O$, $B^{-1} = O^{T}A^{-1}O$. In components, $(B^{-1})_{IJ} = {\bf e}_{I}^{T}A^{-1}{\bf e}_{J}$. This matrix is clearly symmetric and positive definite. Then, the minimal norm $\delta^{2}_{\min}$ can be written as 
\begin{align}
    \delta^{2}_{\min}  = \sum_{IJ=1}^{N}(B^{-1})_{IJ}(\tilde{y}_{I}- {y}_{0}1_{I})(\tilde{y}_{J}- {y}_{0}1_{J})
\end{align}
Now the goal is to find the smallest value of the norm within the domain $\chi^{2}\leq \chi^{2}_{0}$. Due to convexity of $\delta^{2}_{\min}$, the minimum value is attained on the boundary of this domain which satisfies $\chi^{2}= \chi^{2}_{0}$. Thus, we can recast the problem of finding the data point satisfying $\chi^{2}= \chi^{2}_{0}$ which minimizes $\delta^{2}_{\min}$ as a Lagrange multiplier problem:
\begin{align}
 G[\mathbf{y},y_{0},\lambda] = \delta^{2}_{\min}[\mathbf{y},y_{0}] + \lambda (\chi^{2}[\mathbf{y}] - \chi^{2}_{0}) 
\end{align}
We then have the following two equations: $\frac{\partial G}{\partial y_{0}} = \frac{\partial G}{\partial \tilde{y}_{I}} = 0$. These equations simplify to  
\begin{align}
    \sum_{J =1}^{N}(B^{-1})_{IJ}(\tilde{y}_{J} - y_{0}1_{J}) + \frac{\lambda}{\epsilon_{I}^{2}} (\tilde{y}_{I} - \tilde{d}_{I}) = 0 &{\implies} (\tilde{y}_{I} - y_{0}1_{I}) + \lambda\sum_{J =1}^{N}B_{IJ}\frac{(\tilde{y}_{J} - \tilde{d}_{J})}{\epsilon_{J}^{2}}  = 0  \\
    \sum_{I,J =1}^{N}1_{I}(B^{-1})_{IJ}(\tilde{y}_{J} - y_{0}1_{J}) = 0 & {\implies} \sum_{I =1}^{N}\frac{1_{I}(\tilde{y}_{I} - \tilde{d}_{I})} {\epsilon_{I}^{2}} = 0
\end{align}
It is now convenient to define the following variables 
\begin{align}
    \tilde{p}_{J} = \frac{\tilde{d}_{J} - \tilde{y}_{J}}{\epsilon_{J}} \qquad,\qquad
    \tilde{q}_{J} = \frac{\tilde{d}_{J} - y_{0}1_{J}}{\epsilon_{J}} \qquad, \qquad
    M_{IJ} = \frac{B_{IJ}}{\epsilon_{I}\epsilon_{J}}
\end{align}
The Lagrange equations now simplify to 
\begin{align}
    (\tilde{q}_{I} - \tilde{p}_{I}) - \lambda\sum_{J =1}^{N}M_{IJ}\tilde{p}_{J} =0 \\
    \sum_{I=1}^{N} \frac{1_{I}\tilde{p}_{I}}{\epsilon_{I}} =0
\end{align}
In terms of these variables, note that $\delta^{2}_{\min} = \sum_{IJ=1}^{N}M^{-1}_{IJ}(\tilde{q}_{I}-\tilde{p}_{I})(\tilde{q}_{J}-\tilde{p}_{J}) = \lambda^{2}\sum_{I,J=1}^{N}M_{IJ}\tilde{p}_{I}\tilde{p}_{J}$ whereas $\chi^{2} = \sum_{I=1}^{N} \tilde{p}_{I}^{2}$.

Let $\{ \sigma_{r}\}$ and $\{ \textbf{f}_{r}\}$ denote the eigenvalues and eigenvectors of $M_{IJ}$. For $\textbf{p} = \sum_{J=1}^{N} \tilde{p}_J \textbf{e}_J$ and $\textbf{q}= \sum_{J=1}^{N} \tilde{q}_J \textbf{e}_J$, we expand them as follows: $\textbf{p} = \sum_{r=1}^{N} p_{r}\textbf{f}_{r}$ and $\textbf{q} = \sum_{r=1}^{N} q_{r}\textbf{f}_{r}$. This allows us to re-express the first Lagrange equation as $q_{r} - p_{r} - \lambda \sigma_{r}p_{r} = 0$ or $p_{r} = \frac{q_{r}}{1+  \lambda\sigma_{r}}$. All that remains is to solve for $y_0$. We do so by defining two new vectors $\textbf{m}$ and $\textbf{n}$ as $\textbf{m} = \sum_{I=1}^{N}\frac{\tilde{d}_I}{\epsilon_{I}}\textbf{e}_{I}= \sum_{r=1}^{N}m_{r}\textbf{f}_{r}$ and $\textbf{n} = \sum_{I=1}^{N}\frac{1_I}{\epsilon_{I}}\textbf{e}_{I}= \sum_{r=1}^{N}n_{r}\textbf{f}_{r}$. Then notice that $q_{r} = m_{r} - y_{0}n_{r}$ and $p_{r} = \frac{m_{r}}{1+  \lambda\sigma_{r}} - y_{0}\frac{n_{r}}{1+  \lambda\sigma_{r}}$. The second Lagrange equation then implies 
\begin{align}
    0 = \sum_{r}n_{r}p_{r} = \sum_{r}\frac{n_{r}m_{r}}{1+  \lambda\sigma_{r}} - y_{0}\sum_{r}\frac{n_{r}^{2}}{1+  \lambda\sigma_{r}} \nonumber \\
    \implies y_0 = \frac{\sum_{r}\frac{n_{r}m_{r}}{1+  \lambda\sigma_{r}}}{\sum_{r}\frac{n_{r}^{2}}{1+  \lambda\sigma_{r}}}
\end{align}
The only undetermined parameter is the Lagrange multiplier $\lambda$. It is fixed by imposing the condition $\chi^{2} = \sum_{r=1}^{N} p_{r}^{2}(\lambda) = \chi^{2}_{0}$. This is the only step that needs to be performed numerically. Once $\lambda$ has been determined, we can compute the minimal norm as $\delta^{2}_{\min} = \lambda^{2}\sum_{r=1}^{N}\sigma_{r}p^{2}_{r}$. Now, just as in the noiseless case, $\delta^{2}_{\min}$ is really a function of $\alpha$ since the data values $d_{i}\equiv d'_{i;\alpha}$ are functions of $\alpha$. Thus, we need to solve for the value of $\alpha$ numerically which minimizes $\delta^{2}_{\min}$. This provides an estimate of the residue $F(-1)$ in the noisy case.

\section{Appendix C: Double Jackknife Error Estimation for Stabilized Analytic Continuation}

Here we describe the procedure used to estimate the error bars associated with SAC estimates in the regime where only a limited number of data points are available. Our approach is based on a two-fold jackknife resampling scheme applied directly to the raw data. The first resampling layer produces a collection of leave-one-out R\'enyi entropy estimates. For each such estimate, a second resampling layer is used to construct the corresponding covariance matrix. Each R\'enyi entropy estimate, together with its associated covariance matrix, is then passed through the SAC procedure, and the spread of the resulting SAC estimates gives the final jackknife variance estimate. We benchmark this procedure using simulated data, where an independent estimate of the ``true'' variance is available.

We begin with a raw data set
\begin{equation}
    D = \{ X_1, X_2, \cdots, X_{N_B} \},
\end{equation}
where the $X_i$ are taken to be i.i.d. samples. In our setup, the $X_i$ correspond to batched shadow density matrices, where $N_u$ shadows are grouped into $N_B \ll N_u$ batches, averaging within each batch before computing the U-statistics, as defined in~\cite{rathentanglement2023}. We then apply two nested layers of jackknife resampling to this data set.

\begin{enumerate}
    \item \textbf{Outer jackknife:}
    We first construct the leave-one-out data sets
    \begin{equation}
        D_i = \{D\}/X_i .
    \end{equation}
    For each data set $D_i$, we use U-statistics to estimate the desired R\'enyi entropies. We denote the resulting R\'enyi vector, containing the estimated R\'enyi entropies of order $2$ to $k_{\text{max}}$,  by $\mathbf{r}_i$. The outer jackknife therefore produces the collection 
    \begin{equation}
        R = \{\mathbf{r}_1, \mathbf{r}_2, \cdots, \mathbf{r}_{N_B}\}.
    \end{equation}
    of $N_B$ R\'enyi vectors $\mathbf{r}_i$ ($i=1,\dots, N_B$).

    \item \textbf{Inner jackknife:}
    For each outer jackknife data set $D_i$, we perform a second leave-one-out resampling step. This gives
    \begin{equation}
        D_{ij} = \{D_i\}/X_j, \qquad j \neq i .
    \end{equation}
    Thus, for each fixed outer jackknife index $i$, the index $j$ runs over the
remaining $N_B-1$ data points. For every inner jackknife data set $D_{ij}$, we
again use U-statistics to estimate the desired R\'enyi entropies. Suppose our goal is to obtain R\'enyi
entropies up to order $k_{\text{max}}$. Each $D_{ij}$ produces a
$(k_{\text{max}}-1)$-component R\'enyi vector $\mathbf{r}_{ij} (j\neq i)$, corresponding to the R\'enyi orders
$2,\ldots,k_{\text{max}}$. For each fixed $i$, we collect these vectors over all $j\neq i$ and obtain 
an inner-jackknife R\'enyi array of shape
$(N_B-1)\times(k_{\text{max}}-1)$.

    For each fixed $i$, we define the corresponding mean R\'enyi vector
    \begin{equation}
        \tilde{\mathbf{r}}_i = \frac{1}{N_B-1} \sum_{j \neq i}^{N_B} \mathbf{r}_{ij}.
    \end{equation}
    The jackknife covariance matrix associated with the outer sample $D_i$ is then estimated as
    \begin{equation}
        \Sigma_i = \frac{N_B-2}{N_B-1} \sum_{j \neq i}^{N_B}
        (\mathbf{r}_{ij} - \tilde{\mathbf{r}}_i) (\mathbf{r}_{ij} - \tilde{\mathbf{r}}_i)^T .
    \end{equation}
    Repeating this for every outer jackknife sample gives the set of covariance matrices
    \begin{equation}
        \Sigma = \{ \Sigma_1, \Sigma_2, \cdots, \Sigma_{N_B}\}.
    \end{equation}
\end{enumerate}

This construction assigns a covariance matrix $\Sigma_i$ to each outer jackknife R\'enyi vector $\mathbf{r}_i$. We then feed each pair $(\mathbf{r}_i,\Sigma_i)$ into the SAC toolbox, obtaining a corresponding SAC estimate $E_i$. The variance of the SAC estimator is finally obtained by applying the standard outer jackknife variance formula to these estimates:
\begin{equation}
    \text{var} = \frac{N_B-1}{N_B} \sum_{i=1}^{N_B} (E_i - \overline{E})^2,
\end{equation}
where
\begin{equation}
    \overline{E} = \frac{1}{N_B} \sum_{i=1}^{N_B} E_i .
\end{equation}
In this way, the inner jackknife estimates the covariance of the R\'enyi data supplied to SAC, while the outer jackknife estimates the variance of the final nonlinear SAC estimator.

As a proof of concept, we test this double jackknife procedure on simulated data, for which an independent estimate of the ``true'' variance can be obtained from the full ensemble of 1000 simulated experiments. To perform this test, we randomly select a few samples from the 1000 simulated experiments and apply the double jackknife procedure described above. Figure~\ref{fig:Double-Jackknife variance performance} shows a representative result from one such sample. It compares the standard deviation obtained by taking the square root of the double jackknife variance estimate with the corresponding ``true'' standard deviation computed from the full ensemble of 1000 simulated data sets, i.e., the same ensemble standard deviation used to assign the error bars in Fig.~3 of the main text.

\renewcommand{\thefigure}{S\arabic{figure}}
\begin{figure}
    \centering
    \includegraphics[width=0.8\linewidth]{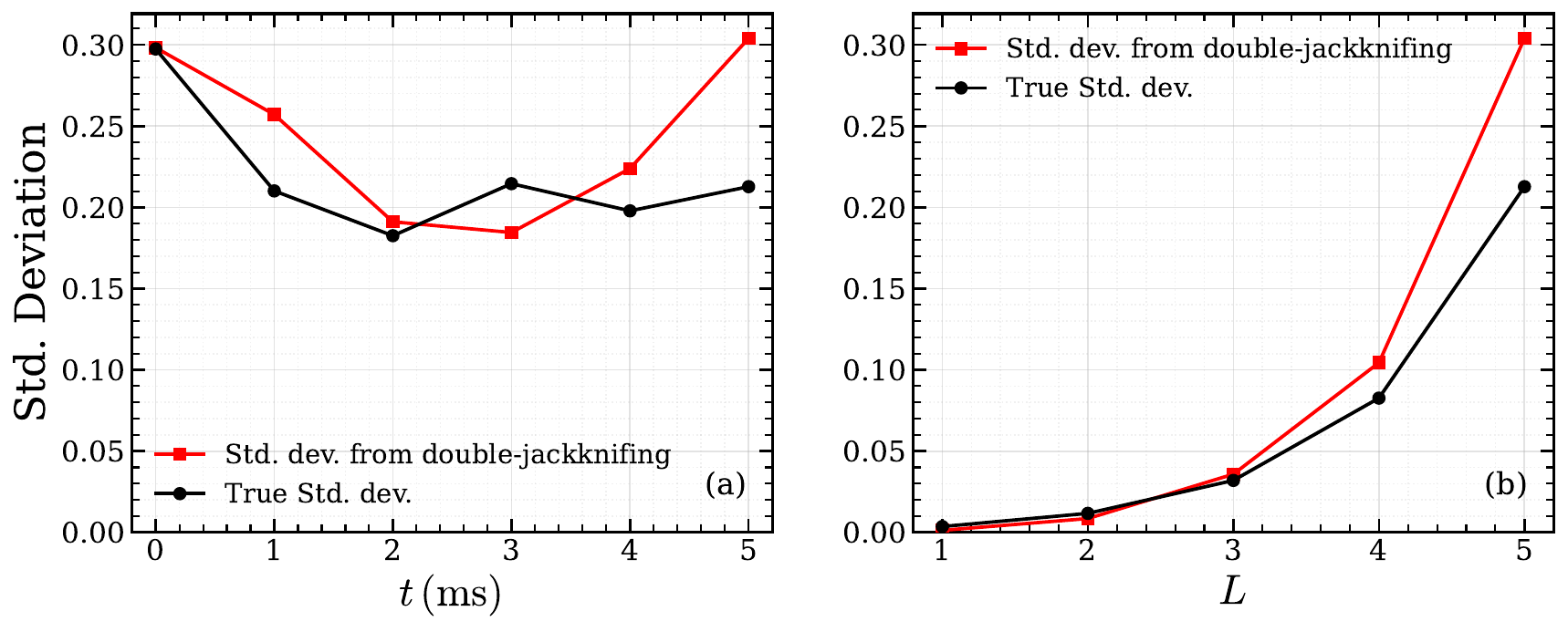}
    \caption{(a): Standard deviation $\sigma$ = $\sqrt{\text{var}}$ obtained using double-jackknifing on a randomly chosen simulation sample vs. true standard deviation plotted at different times for subsystem $L = 5$. (b): $\sigma$ plotted for different subsystems at $t = 5$ms.}
    \label{fig:Double-Jackknife variance performance}
\end{figure}

\noindent\textit{Remark.} 
In the procedure described above, the inputs to SAC are the outer jackknife R\'enyi vectors $R$ together with their associated covariance matrices $\Sigma$. An alternative choice is to use the jackknife-corrected R\'enyi vectors instead. We denote this corrected set by $R^{JK}$, defined as
\begin{equation}
    R^{JK}
    =
    \left\{
    \mathbf{r}_i^{JK}
    \ \bigg|\
    \mathbf{r}_i^{JK}
    =
    (N_B-1) \mathbf{r}_i
    -
    \frac{N_B-2}{N_B-1}
    \sum_{j \neq i} \mathbf{r}_{ij}
    \right\}.
\end{equation}
One may then feed the pairs $(\mathbf{r}_i^{JK},\Sigma_i)$, rather than $(\mathbf{r}_i,\Sigma_i)$, into the SAC procedure. For the data sets considered here, we find that the jackknife corrections to the R\'enyi vectors are numerically small. Consequently, the corrected set $R^{JK}$ is very close to $R$, so the SAC estimates obtained from the two choices are nearly identical. The corresponding jackknife variance estimates, and hence the resulting SAC error bars, are therefore effectively unchanged.

\end{document}